\documentclass[%
reprint,
superscriptaddress,
showpacs,preprintnumbers,
amsmath,amssymb,
aps,
pre,
]{revtex4-1}
\usepackage{dcolumn}% Align table columns on decimal point
\usepackage{bm}% bold math

\usepackage{graphicx}
\newcommand{\vect}[1]{{\ensuremath{\boldsymbol{#1}}}}

\begin{document} 

  \title{Reactivity Boundaries to Separate the Fate of a Chemical Reaction Associated with Multiple Saddles}
  
  \author{Yutaka Nagahata}
    %\email{yutaka\_nagahata@mail.sci.hokudai.ac.jp}
    \affiliation{Graduate School of Life Science, Hokkaido University, Kita 12, Nishi 6,Kita-ku, Sapporo 060-0812, Japan}
  \author{Hiroshi Teramoto}
    %\email{teramoto@es.hokudai.ac.jp}  
    \affiliation{Graduate School of Life Science, Hokkaido University, Kita 12, Nishi 6,Kita-ku, Sapporo 060-0812, Japan}
    \affiliation{Molecule and Life Nonlinear Sciences Laboratory, Research Institute for Electronic Science, Hokkaido University, Kita 20 Nishi 10, Kita-ku, Sapporo 001-0020, Japan}
  \author{Chun-Biu Li}
    %\email{cbli@es.hokudai.ac.jp}
    \affiliation{Molecule and Life Nonlinear Sciences Laboratory, Research Institute for Electronic Science, Hokkaido University, Kita 20 Nishi 10, Kita-ku, Sapporo 001-0020, Japan}
    \affiliation{Graduate School of Science, Department of Mathematics, Hokkaido University, Kita 12, Nishi 6,Kita-ku, Sapporo 060-0812, Japan}
    \affiliation{Research Center for Integrative Mathematics, Hokkaido University, Kita 20, Nishi 10, Kita-Ku, Sapporo, Hokkaido, 001-0020, Japan}
  \author{Shinnosuke Kawai}
    %\email{skawai@es.hokudai.ac.jp}
    \affiliation{Graduate School of Life Science, Hokkaido University, Kita 12, Nishi 6,Kita-ku, Sapporo 060-0812, Japan}
    \affiliation{Molecule and Life Nonlinear Sciences Laboratory, Research Institute for Electronic Science, Hokkaido University, Kita 20 Nishi 10, Kita-ku, Sapporo 001-0020, Japan}
  \author{Tamiki Komatsuzaki}
    \email{tamiki@es.hokudai.ac.jp}
    \affiliation{Graduate School of Life Science, Hokkaido University, Kita 12, Nishi 6,Kita-ku, Sapporo 060-0812, Japan}
    \affiliation{Molecule and Life Nonlinear Sciences Laboratory, Research Institute for Electronic Science, Hokkaido University, Kita 20 Nishi 10, Kita-ku, Sapporo 001-0020, Japan}
    \affiliation{Research Center for Integrative Mathematics, Hokkaido University, Kita 20, Nishi 10, Kita-Ku, Sapporo, Hokkaido, 001-0020, Japan}
  \date{\today}
  
\begin{abstract}
  Reactivity boundaries that divide the origin and destination of
  trajectories are crucial of importance to reveal the mechanism of
  reactions, which was recently found to exist robustly even at high
  energies for index-one saddles [Phys. Rev. Lett. {\bf 105}, 048304 (2010)].
  Here we revisit the concept of the reactivity boundary and propose a
  more general definition that can involve 
  a single reaction %(i.e., not a sequence of different reactions)
  associated with a bottleneck made up of higher index saddles 
  and/or several saddle points with different indices,
  where the normal form theory, based on expansion around a single stationary point, does not work.
  We numerically demonstrate the reactivity
  boundary by using a reduced model system of the $\mathrm{H}_5^+$ cation
  where the proton exchange reaction takes place through a
  bottleneck made up of two index-two saddle points and two index-one saddle points.
  The cross section of the reactivity boundary in the reactant region of
  the phase space reveals which initial conditions are effective in making
  the reaction happen, and thus sheds light on the reaction mechanism.
\end{abstract}
  
  %\keywords{}
  
  \maketitle
  
  %\section{}
  %\subsection{}
  %\subsubsection{}
 
\section{Introduction}

  Studies of chemical reaction dynamics aims for understanding of how and why a system 
  proceeds from its initial state to the final state in the process of reaction. 
  Special interest lies in the question of what initial conditions make the reaction happen.
  Classically, the process of chemical reaction can be regarded as motion of a point
  in the phase space propagating from a region corresponding to the reactant to another region corresponding to the product.
  Some phase space points in the reactant region may go into the product region after time propagation,
  whereas other phase space points stay in the reactant region without undergoing the reaction.
  In between these reactive initial conditions and non-reactive ones lies a boundary
  which we simply call here reactivity boundary %\cite{Kawai2010a}
  \
  that was previously described by various words, such as 
  ``\textit{boundary trajectories}''
  \cite{Pechukas1976,Pechukas1977,Sverdlik1978,Pollak1978,Pechukas1979,Pollak1979,Child1980,Pollak1980,Pollak1980a,Pollak1980b,Pechukas1981,Pollak1981,Pollak1981a,Pollak1981b}
  asymptotic to periodic orbit dividing surface (pods)
  \cite{                                                                          Child1980,Pollak1980,Pollak1980a,Pollak1980b,Pechukas1981,Pollak1981,Pollak1981a,Pollak1981b},
  ``\textit{boundary of}''
  \cite{Pechukas1976,Pechukas1977,Sverdlik1978,Pollak1978,Pechukas1979,Pollak1979,Child1980,Pollak1980,Pollak1980a,Pollak1980b,Pechukas1981,Pollak1981,Pollak1981a,Pollak1981b}
  reactivity bands\cite{Wall1958,Wall1961,Wall1963,Wright1975,Wright1976,Wright1977,Laidler1977,Tan1977,Wright1978,Andrews1984,Grice1987},
  ``\textit{tube}''\cite{DeAlmeida1990}, ``\textit{cylindrical manifold}''\cite{DeAlmeida1990},
  ``\textit{impenetrable barriers}''\cite{Wiggins2001},
  ``\textit{stable/unstable manifold}'' of normally hyperbolic invariant manifolds (NHIM)\cite{Wiggins2001},
  ``\textit{reaction boundaries}''\cite{Kawai2010a},
  and also described on certain sections, such as
  ``\textit{reactivity bands}''\cite{Wall1958,Wall1961,Wall1963,Wright1975,Wright1976,Wright1977,Laidler1977,Tan1977,Wright1978,Andrews1984,Grice1987},
  ``\textit{reactivity map}''  \cite{                           Wright1975,Wright1976,Wright1977,Laidler1977,Tan1977,Wright1978,Andrews1984,Grice1987},
  ``\textit{reactive island}''\cite{DeAlmeida1990}.
  The general definition
  %and extraction 
  of the reactivity boundary is the main subject of this paper.
  
  The reactivity boundary is often discussed in relation to saddle points.
  A saddle point on a multi-dimensional potential energy surface is defined as a stationary point at which the Hessian matrix does not have zero eigenvalues and, at least, one of the eigenvalues is negative. 
  Saddle points are classified by the number of the negative eigenvalues and a saddle that has $n$ negative eigenvalues is called an \textit{index-$n$ saddle}.
  Especially the index-one saddle on a potential surface has long been
  considered to make bottleneck of reactions\cite{Glasstone1941,Steinfeld1989},
  with the sole unstable direction corresponding to the ``reaction coordinate.''
  This is because index-one saddle is considered to be the lowest
  energy stationary point connecting two potential minima, of
  which one corresponds to the reactant and the other to the product,
  and the system must traverse the vicinity of the index-one saddle from the reactant to the product \cite{Zhang2006,Skodje2000,Shiu2004}.

  Such reactivity boundaries have been investigated from early period of the study of reaction dynamics. 
  Especially reactivity boundaries of atom-diatom reactions were extensively studied by
  Wright et al.\cite{Wright1975,Wright1976,Wright1977,Tan1977,Wright1978,Laidler1977} and 
  Pechukas et al.\cite{Pollak1980a,Pollak1981b,Pollak1981,Pechukas1976,Sverdlik1978,Pollak1979,Pollak1980b,Child1980,Pollak1978,Pollak1981a,Pollak1980,Pechukas1979,Pechukas1977,Pechukas1981}. 
  At very early period, Wigner introduced asymptotic reactant and product regions to calculate reaction rate in the line of his achievement of the transition state theory\cite{Wigner1937}. 
  Independently, Wright et al. showed reactive bands, which had been found by Wall et al.\cite{Wall1961,Wall1963,Wall1958}, in the reactivity maps of $\mathrm{H}+\mathrm{H}_2$ and its isotopic variants\cite{Wright1975,Wright1976,Wright1977,Tan1977,Wright1978,Laidler1977} that consist of bands of nonreactive regions and reactive regions of each product. 
  The approach was initiated by Ref.~\onlinecite{Wright1975} to see the
  origin of continuous shift of peak in graph of initial relative
  translational (kinetic) energy versus time spent in {``\it{reaction shell}''}
  for given initial vibrational phase angles. After
  Ref.~\onlinecite{Wright1975}, a series of study was 
  reported for collinear (1D)\cite{Wright1976}, isotope\cite{Wright1977}, coplanar (2D)\cite{Tan1977} reactions and 3D\cite{Wright1978} reaction and also on an improved potential energy surface\cite{Laidler1977} with the plot of initial relative translational energy versus initial phase angle $\theta$. 
  Chesnavich et al. observed boundary trajectories of the 
  collision-induced dissociation of $\mathrm{H}+\mathrm{H}_2$ 
  reaction\cite{Andrews1984,Grice1987} that divide reactive ($\mathrm{H}_2+\mathrm{H}$), non-reactive ($\mathrm{H}+\mathrm{H}_2$) and dissociative ($\mathrm{H}+\mathrm{H}+\mathrm{H}$) regions in phase space.

  Pechukas et al. revealed the role of periodic orbit dividing surface
  in two-dimensional collinear atom-diatom reaction systems
  \cite{Pollak1980a,Pollak1981b,Pollak1981,Pechukas1976,Sverdlik1978,Pollak1979,Pollak1980b,Child1980,Pollak1978,Pollak1981a,Pollak1980,Pechukas1979,Pechukas1977,Pechukas1981}.
  The importance of periodic orbit around interaction region was first recognized by Pechukas\cite{Pechukas1976}. 
  The series of research can be described by his words at very beginning.
  \begin{quote}
  Somewhere between these two trajectories is a ``dividing'' trajectory that falls away, neither to reactant nor to product; this is the required ``vibration,'' across the saddle point region but not necessarily through the saddle point, and the curve executed on the plane by the vibration is the best transition state at that energy.
  \end{quote}
  Pechukas and Pollak\cite{Pechukas1977} and Sverdlik and Koeppl\cite{Sverdlik1978} started to observe such trajectories in the region of index-one saddles of two dimensional systems and recognized as {``\it{unstable invariant classical manifold}''}\cite{Pechukas1981} and call them periodic orbit dividing surface (pods)\cite{Pollak1980}. 
  The pods can be identified as the best transition
  state\cite{Pechukas1979} when there is only one pods at given
  energy. Pechukas and Pollak investigated the advantage of pods against
  variational TST\cite{Pollak1978} and unified statistical
  theory\cite{Pollak1979}. They also revealed its role in the application of statistical theories to reaction dynamics\cite{Pollak1979,Pollak1980a,Child1980} and provided an iterative method to calculate reaction probability\cite{Pollak1980b}. 
  After the series of classical investigation they started to look at
  adiabatic motion perpendicular to pods\cite{Pollak1981a} and quantum
  correspondence\cite{Pollak1981} and experimental
  correspondence\cite{Pollak1981b} were elucidated. 
  Those studies were mostly done on two degrees of freedom (DoFs) 
  systems. The problem one of high dimension in the region of
  index-one saddles was later overcome \cite{Komatsuzaki1996,Komatsuzaki1997,Komatsuzaki1999,Komatsuzaki2001,Wiggins2001,UzerNonlin02,Bartsch2005a,Li2006,Kawai2010a,Hernandez2010,NFLrev,QNFrev,QTDNF,NFrotSK,NFrotUC,Teramoto2011,book_adv05,book_adv11,Jaffe2005,Martens2002,Komatsuzaki2003,DeAlmeida1990}.

  The dynamics around the saddle point is recently investigated extensively 
  in terms of nonlinear dynamics \cite{Komatsuzaki1999,Komatsuzaki2001,Wiggins2001,UzerNonlin02,Bartsch2005a,Li2006,Kawai2010a,Hernandez2010,NFLrev,QNFrev,QTDNF,NFrotSK,NFrotUC,Teramoto2011,book_adv05,book_adv11,Jaffe2005,Martens2002,Komatsuzaki2003,DeAlmeida1990},
  in the context of transition state (TS) theory \cite{Glasstone1941,Steinfeld1989} in molecular science.
  Among them, particularly relevant to the present work is the finding of the ``tube''\cite{DeAlmeida1990} structures in phase space to conduct the reacting trajectories from the reactant to the product across an index-one saddle.
  These studies revealed the firm theoretical ground for the robust
  existence of the reactivity boundaries emanating from the saddle region 
  as well as the no-return TS in the phase space\cite{book_adv05,book_adv11}.
  The scope of the dynamical reaction theory\cite{book_adv11} is not limited to only chemical reactions, but also includes,
  for example, ionization of a hydrogen atom under electromagnetic
  fields \cite{Wiggins2001,UzerNonlin02}, isomerization of clusters
  \cite{Komatsuzaki1999,Komatsuzaki2001}, orbit designs in solar systems \cite{Jaffe2002}, and so forth.
  Recently, these approaches have been generalized to dissipative
  multidimensional Langevin equations
  \cite{Bartsch2005a,Hernandez2010,NFLrev}, based on a seminal work by Martens\cite{Martens2002}, laser-controlled chemical
  reactions with quantum effects \cite{QNFrev,QTDNF}, systems with
  rovibrational couplings \cite{NFrotSK,NFrotUC}, and showed the robust
  existence of reaction boundaries even while a no-return TS ceases to
  exist \cite{Kawai2010a}.

  For complex systems, the potential energy surface becomes more complicated, and transitions from a potential basin to another
  involve not only index-one saddles but also higher index saddles \cite{Shida2005,Sicardy2010,Minyaev1997,Minyaev2004,Getmanskii2008,Huang2006,Shank2009,Xie2005,Bowman2011}. 
  Recently the role of index-two saddle was revealed several dynamical aspects.
  For example, a simulation study on ``phase transitions'' from solid-like phase to liquid-like phase in a seven-atomic cluster \cite{Shida2005}
  showed that trajectories spend more time in the region of higher index saddles as the total energy of the system increases.
  Under the onset of ``melting'', its occupation ratio around the index-two saddles correlates to its Lindemann's $\delta$ and the configuration entropy that are well-known indices of phase transition.
  Another example is a systematical survey of global stability of
  the triangular Lagrange points L4 and L5 under the condition that
  the secondary mass $\mu$ is larger than the Gascheau's value $\mu_G$ 
  (also known as the Routh value) in the restricted planar circular three-body problem \cite{Sicardy2010}. 
  Those Lagrange points become index-two saddle points when the condition
  $\mu>\mu_G$ is met, and the range of $\mu$ was identified where the Lagrange points have global stability and periodic stable orbits around them.

  Chemical reactions associated with index-two saddles were also reported in several molecular systems \cite{Minyaev1997,Minyaev2004,Getmanskii2008,Huang2006,Shank2009,Xie2005,Bowman2011} 
  by using several searching algorithms (section 6.3 p.~298 of Ref.~\onlinecite{Wales2004} and references therein). 
  However index-two saddles have got less interests than index-one saddles.
  This may be because of the Murrell-Laidler theorem \cite{Murrell1968}
  that states the minimum energy path does not pass through any index-two saddle points.
  However one can still find many studies such as 
  aminoborane\cite{Minyaev1997}, $\text{PF}_3$\cite{Minyaev1997}, $\text{NH}_5$\cite{Minyaev2004}, $\text{NF}_2\text{H}_3$\cite{Getmanskii2008}, 
  water dimer\cite{Huang2006,Shank2009}, $\text{H}_5^+$\cite{Xie2005}, $\text{H}_2\text{CO}$\cite{Bowman2011} 
  that identify a variety of index-two saddles in molecular isomerization reactions. 
  
  Significant difference between reactions associated with a
  bottleneck made of an index-one saddle and those through higher index
  saddle is that a single higher index saddle does
  not necessarily serve as a bottleneck from one potential basin
  to another since index-$n$ $(>1)$ saddles are almost always
  accompanied with saddles of index less than $n$.
  Therefore, reactions associated with higher index saddle(s) are
  dominated  by a bottleneck made up of multiple saddles, and so are its phase space structures.
  This non-local property of the bottleneck 
  %in an elementary,  single reaction 
  is an essential difficulty in treating a reaction associated with higher index saddles.

  To reveal the fundamental mechanism of the passage through a saddle
  with index greater than one, the phase space structure was recently
  studied on the basis of normal form (NF) theory \cite{Ezra2009a,Collins2011,Haller2010,Haller2011}. 
  For example the pioneering studies to extend the dynamical reaction theory into
  higher index saddles were reported \cite{Ezra2009a} for concerted reactions.
  A dividing surface to separate the reactant and the product was proposed for higher index saddles \cite{Collins2011} 
  and the associated phase space structure was also discussed \cite{Haller2010,Haller2011}.
  Those studies are based on NF theory, and therefore relies on two assumptions: 
  One is that no linear ``resonance'' is postulated between more
  than one reactive modes and the other is that the local dynamics
  around the index-two or higher index saddle plays a dominant role in
  determining the destination of the trajectory. 
  For the former assumption, Toda\cite{Toda2008} addressed that linear resonance 
  between two reactive modes may introduce breakdown of the reactivity boundary. 
  As for the latter assumption,
  Nagahata et al.\cite{Nagahata2013a} reported recently that
  the reactivity boundary extracted by normal form does not
  necessarily give the barrier separating the reactivity % and nonreactive regions 
  in the original coordinate space for higher index saddles.
  %especially along the less repulsive degrees of freedom.
  Moreover, as described above, an %an
  index-two saddle often coexists with index-one saddles
  and therefore the reaction dynamics or the ``bottleneck'' should be determined through interplay among multiple saddle points.
  Additionally, current theory for invariant manifold that may dominate reactions associated with an index-two saddle and a higher index saddle 
  are only for the largest repulsive direction\cite{Haller2010,Haller2011}.
  However, for example, Minyaev et al. \cite{Minyaev1997} showed that aminoborane 
  has internal rotation associated with an index-two saddle and
  index-one saddles, and that
  the weaker repulsive direction around the index-two saddle,
  corresponding to the hindered internal rotation, connects the two minima.

  Most studies for reaction associated with higher index saddles are based on NF, a perturbation theory around a single stationary point, and assume that NF can capture those reaction dynamics.
  To validate those studies, however, it is needed first to clarify the concept of reactivity boundaries in reactions associated with a bottleneck made up of multiple saddles.
  The reactivity map\cite{Wright1975,Wright1976,Wright1977,Laidler1977,Tan1977,Wright1978,Andrews1984,Grice1987} and Pechukas' foresight\cite{Pechukas1976} are still important to generalize the concept to make it applicable when the reaction dynamics is not dominated by a single saddle point.

  In the present paper, we first review the concept of reactivity
  boundaries for the linear system in Sec.\ref{ssec:NM}. Then we generalize the concept to the reactions associated with a bottleneck possibly made up of multiple saddle points in Sec\ref{ssec:RB}.
  In Sec.\ref{sec:demo} 
  we demonstrate the numerical extraction of reactivity boundaries in a chemical system with a bottleneck made up of multiple saddle points
  including both index-one and index-two saddles.
  The investigation reveals what initial condition should be prepared to make the reaction happen,
  and why such initial conditions lead to reactions.

\section{Theory}
  \label{sec:th}
  In this section, we revisit the concept of reactivity boundaries
  developed previously (Sec.~\ref{ssec:NM}) and propose a more general
  definition that can involve a single reaction
  %(i.e., not a sequence of different reactions)
  associated with a bottleneck made up of higher index saddles and/or several saddle points with different indices, where the normal form theory, based on expansion around a single stationary point, does not work (Sec.~\ref{ssec:RB}).
  \subsection{Linearized Hamiltonian}%Normal mode}
\begin{figure}
      \includegraphics[width=6.5cm,bb=0 0 195 187]{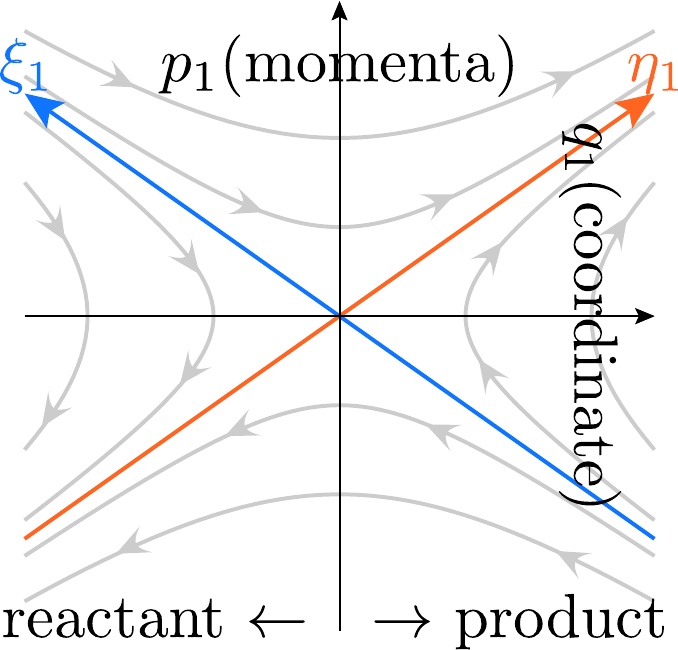}
      \caption{
        (color online).
        Phase space flow of the normal
        mode with negative curvature (hyperbolic degree of
        freedom). Reactant and product are defined
        by the sign of $q_1$.
        $\eta_1 = 0$(or  $\xi_1$-axis) divides the destination of trajectories;
        Trajectories in $\eta_1>0$ go into the product side ($q_1>0$) as $t\rightarrow+\infty$
        and those in $\eta_1<0$ go into the reactant side ($q_1<0$). 
        Similarly, $\xi_1  = 0$(or $\eta_1$-axis) divides the origin of trajectories;
        Trajectories in $\xi_1>0$ originate from the reactant side
        and those in $\eta_1<0$ from the product side.
        }
    \label{fig:fig1}
    \end{figure}

    \label{ssec:NM}
    In this subsection we review the concept of reactivity boundaries developed previously based on the theory of dynamical systems
    \cite{Komatsuzaki1999,Komatsuzaki2001,Wiggins2001,UzerNonlin02,Bartsch2005a,Li2006,Kawai2010a,Hernandez2010,NFLrev,QNFrev,QTDNF,NFrotSK,NFrotUC,Teramoto2011,book_adv05,book_adv11,Jaffe2005,Martens2002,Komatsuzaki2003,DeAlmeida1990}.
    One of the simplest example of reactivity boundaries can be seen in the normal mode (NM) approximation.
    If the total energy of the system is just slightly above a stationary point, the $n$-DoFs Hamiltonian $H$ can well be approximated by a NM Hamiltonian $H_0$
    \begin{equation} \label{eq:nm}
      H(\vect{p},\vect{q})
      \approx 
      H_{0}(\vect{p},\vect{q})
      =
      \sum_{j=1}^n \frac{1}{2}(p_j^2+k_j q_j^2)
    \end{equation}
    with NM coordinates 
    \vect{q}=$(q_1,\dots,q_n)$ 
    and their conjugate momenta
    \vect{p}=$(p_1,\dots,p_n)$, 
    where $k_j \in \mathbb{R}$ is the ``spring constant'' or 
    the curvature of the potential energy surface along the $j$th direction. 
    The constants $k_j$ can be positive or negative.
    If $k_j<0$, the potential energy is maximum along the $j$th direction.
    Then the direction exhibits an unstable motion corresponding to ``sliding down the barrier,'' 
    and can be regarded as ``reaction coordinate.''
    The index of the saddle corresponds to the number of negative $k_j$'s.
    Phase space flow of the DoF with negative $k_j$ is depicted in Fig.~\ref{fig:fig1}. 
    Here one can introduce the following coordinates
    \begin{eqnarray} \label{eq:xieta_nm}
      \eta_j=& (p_j+\lambda_j q_j)/(\lambda_j\sqrt{2})
      , ~~~%\cr
      \xi_j =&(p_j-\lambda_j q_j)/\sqrt{2}
      ,
    \end{eqnarray}
    corresponding to eigenvectors of the coefficient matrix of
    the linear differential equation (Eq.~\ref{eq:nm}) with eigenvalue $\lambda_j=\pm\sqrt{-k_j}$.
    Here one can also introduce another set of coordinates
    \begin{eqnarray} \label{eq:Itheta_nm}
      I_j=& \xi_j \eta_j
      , ~~~%\cr
      \theta_j =& \ln|\lambda_j\eta_j/\xi_j|/2
      ,
    \end{eqnarray}
    called ``action'' and ``angle'' variables.
    When Eq.\;\eqref{eq:nm} holds, the action variable is an integral of motion, and trajectories run along the hyperbolas given by $I_j=\mathrm{const.}$ shown by gray lines in Fig.~\ref{fig:fig1}.
    The $\eta_j$- and $\xi_j$-axes run along the asymptotic lines of the hyperbolas in Fig.~\ref{fig:fig1}.
    The Hamiltonian equation of motion can be written as
    \begin{equation} \label{eq:NMeqm}
      \dot{\vect{\zeta}}_j\approx-L_{H_0} \vect{\zeta}_j = -\lambda_j L_{I_j} \vect{\zeta}_j,
      = \begin{pmatrix}
      -\lambda_j & 0 \cr 
      0 & \lambda_j 
      \end{pmatrix} 
      \vect{\zeta}_j,
    \end{equation}
    where $\vect{\zeta}_j=(\xi_j,\eta_j)^\mathrm{T}$, and the Lie derivative $L_F$ is defined as
    $
      L_F \vect{\zeta}_k = \{F,\vect{\zeta}_k\} 
      = \sum_{j=1}^n 
       \frac{\partial F}{\partial \eta_j}\frac{\partial \vect{\zeta}_k}{\partial  \xi_j}
      -\frac{\partial F}{\partial  \xi_j}\frac{\partial \vect{\zeta}_k}{\partial \eta_j}
    $.
    One can tell the destination region and the origin region of trajectories 
    from the signs of $\eta_j$ and $\xi_j$ as follows:
    If $\eta_j>0$, the trajectory goes into $q_j>0$ and if $\eta_j<0$, then the trajectory goes into $q_j<0$.
    Therefore one can tell the destination of trajectories from the sign of $\eta_j$.
    Similarly, the origin of trajectories can be told from the sign of $\xi_j$.
    Hereafter we call the set $\eta_j=0$ ``destination-dividing set,'' and
    $\xi_j=0$ ``origin-dividing set,'' and each of these two sets constitute ``reactivity boundaries''.

    When the NM picture dominates the dynamics around the stationary point, the
    form of Eq.~\eqref{eq:NMeqm} enables us to identify the fate of reaction.
    This is also generally the case if one can achieve a canonical transformation 
    to turn the Hamiltonian into the form of $H=H(\vect{I})$,
    even though $\lambda_j$s are depends on initial ${I_j}$s
    .
    %in the same manner with the NM case explained in Sec.~\ref{ssec:NM},and 
    This transformation has been mostly achieved by
    the normal form theory based on expansion around a single stationary point. The theory has been applied and developed to
    elucidate the mechanism of several reaction dynamics about a decade
    \cite{Komatsuzaki1999,Komatsuzaki2001,Wiggins2001,UzerNonlin02,Bartsch2005a,Li2006,Kawai2010a,Hernandez2010,NFLrev,QNFrev,QTDNF,NFrotSK,NFrotUC,Teramoto2011,book_adv05,book_adv11}.
    For practical applications the Lie
    canonical perturbation theory, %originally 
    developed by a Japanese astrophysicist
    Gen-Ichiro Hori\cite{Hori1966,Hori1967}
    (and equivalent theory was independently developed by Deprit%that is equivalent to Deprit's perturbation theory
    \cite{Campbell1970,Deprit1969a}), 
    has been frequently used.

  \subsection{Reactivity boundary}
    \label{ssec:RB}
    For complex molecular systems, the potential energy
    surface becomes more complicated, and a single transition from
    a potential basin to another involves not only index-one
    saddles but also higher index saddles.
    The normal form theory shown in Sec.~\ref{ssec:NM}, % and \ref{ssec:NF}, 
    based on expansion around a single stationary point, 
    may not work well for such complex systems, where the
    fate of the reaction may not be
    dominated solely by the local property of the potential around the point.
    Therefore the definition of the reactivity boundaries should not be based on perturbation theory.
    In this subsection, we seek for a more general definition of reactivity boundaries, 
    so that the definition can describe invariant objects previously studied (such as impenetrable barriers\cite{Wiggins2001} and reactive island\cite{DeAlmeida1990}),
    to analyze more complicated reactions by following the Pechukas' foresight\cite{Pechukas1976}.
 
\begin{figure}
      \includegraphics[width=8.5cm,bb=0 0 280 99]{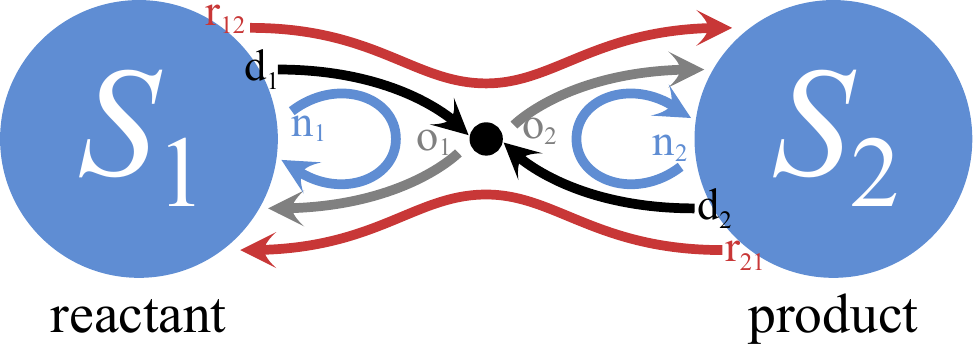}
        \caption{
        (color online).
        Blue large circles represent states $S_1$ and $S_2$.
        Arrows represent particular sorts of trajectories;
        blue arrows ($\mathrm{n}_1$ and $\mathrm{n}_2$) represent non-reactive trajectories,
        while red ones ($\mathrm{r}_{12}$ and
        $\mathrm{r}_{21}$) represent reactive trajectories.
        Black arrows ($\mathrm{d}_1$ and $\mathrm{d}_2$) and
        gray arrows ($\mathrm{o}_1$ and $\mathrm{o}_2$)
        represent trajectories in the destination-dividing set,
        and those in the origin-dividing set, respectively.
        }
      \label{fig:fig2}
    \end{figure}

    A ``state,'' which may refer to reactant or product, forms a certain region 
    in the phase space $\Omega$. Let us denote the states by $S_1, \dots, S_N$, which are disjoint subsets of $\Omega$ ($S_j \subset \Omega$ and $S_i \cap S_j = \emptyset$ where $i,j=1,2,\dots,N$ and $i \neq j$). % and $S_2$.
    %These regions should not have an overlap in the phase space. 
    In between the regions corresponding to the states, there can be intermediate an region $\Omega_0$
    that do not belong to any of the states (Fig.~\ref{fig:fig2}): $\Omega = S_1 \sqcup \dots \sqcup S_N \sqcup \Omega_0$.
    Most of the trajectories in the intermediate region eventually go into
    either of the states 
    as time proceeds. Likewise, when propagated backward in
    time, most of them turn out to originate from either of the states.
    Consider a set of trajectories that originate from the state
    $S_1$ and go into the state $S_2$ ($\mathrm{r}_{12}$ in Fig.~\ref{fig:fig2}), 
    and another set consisting of trajectories that originate from $S_1$
    and go back into the same state $S_1$ ($\mathrm{n}_1$ in Fig.~\ref{fig:fig2}).
    In between these two sets of trajectories there may lie a boundary which consists of trajectories
    that do not go into either of the states ($\mathrm{d}_1$ in Fig.~\ref{fig:fig2}). 
    In the cases discussed in Sec.~\ref{ssec:NM}, such trajectories
    were seen to asymptotically approach into some invariant set(s) in the intermediate region.
    Suppose there exist such an invariant set $\Omega_S$, which is a co-dimension two subset of $\Omega_0$.
    We then consider co-dimension one subset $\Omega_{OD}, \Omega_{DD} \subset \Omega_0$ satisfying $\lim_{T\rightarrow \infty} \phi^T(\Omega_{OD})=\Omega_S$ and $\lim_{T \rightarrow -\infty} \phi^T(\Omega_{DD})=\Omega_S$ as follows:
    %
    %Let us thus consider the following two sets.
    %
    \begin{itemize}
      \item{Destination-dividing set $\Omega_{DD}$ ($\mathrm{d}_1$ and $\mathrm{d}_2$ in Fig.~\ref{fig:fig2})}\\
      A set of trajectories whose origin belongs to a certain state but whose destination does not belong to any state.  
      \item{Origin-dividing set  $\Omega_{OD}$ ($\mathrm{o}_1$ and $\mathrm{o}_2$ in Fig.~\ref{fig:fig2})}\\
      A set of trajectories whose destination belongs to a certain state but whose origin does not belong to any state.  
    \end{itemize}
    The former set constitutes a boundary dividing the destination regions of trajectories, 
    whereas the latter constitutes a boundary dividing the origin regions of trajectories. 
    The set of the trajectories (invariant
    set) that satisfy one of the above conditions will be called reactivity boundary in the following.
    The asymptotic limit $\Omega_S$ of the reactivity boundary, which belongs to neither reactant nor product, will be called ``seed'' of reactivity boundaries.
    The definition of the reactivity boundary (the destination- or the origin-dividing set) can apply systems with multiple states, since the definition of the reactivity boundaries are only based on a single state. 
    This definition of the reactivity boundaries and their seed is 
    a generalization of
    the previous invariant objects (the stable and unstable
    manifolds of NHIM, and the NHIM, respectively)
    studied in literature \cite{Komatsuzaki1999,Komatsuzaki2001,Wiggins2001,UzerNonlin02,Bartsch2005a,Li2006,Kawai2010a,Hernandez2010,NFLrev,QNFrev,QTDNF,NFrotSK,NFrotUC,Teramoto2011,book_adv05,book_adv11}
    and summarized in Sec.~\ref{ssec:NM}.% and \ref{ssec:NF}.
 
\section{Numerical Demonstrations}
  \label{sec:demo}

  \subsection{Three DoFs model of $\text{H}_5^+$}

    We demonstrate here a numerical calculation of the reactivity boundary 
    defined in Sec.~\ref{sec:th}
    with a model $\mathrm{H}_5^+$ system.
    This cation plays an important role in interstellar chemistry,
    especially because of the proton exchange reaction 
    $\mathrm{H}_3^+ + \mathrm{HD} \rightleftharpoons \mathrm{H}_2 + \mathrm{H}_2\mathrm{D}^+$
    occurring through the $\mathrm{H}_4\mathrm{D}^+$ intermediate.
    As shown in the previous \textit{ab initio} calculation \cite{Xie2005}, 
    the most stable structure of the $\mathrm{H}_5^+$ system
    is a weakly bound cluster of $\mathrm{H}_2$ and $\mathrm{H}_3^+$ moieties,
    with the $\mathrm{H}_2$ standing perpendicular to the $\mathrm{H}_3^+$ molecular plane. 
    Being a multi-body system, the $\mathrm{H}_5^+$ cation undergoes various isomerization reactions.
    Taking the four lowest stationary points 
    (one minimum, two index-one saddle points, and one index-two saddle),
    we have two reaction directions. One is a torsional isomerization where the $\mathrm{H}_2$ flips by 180$^\circ$
    with the planar structure corresponding to the saddle point.
    The other is the proton exchange between the two moieties 
    $\mathrm{H}_2 + \mathrm{H}_3^+ \rightleftharpoons \mathrm{H}_3^+ + \mathrm{H}_2$. 
    
      \begin{figure}
        \includegraphics[width=8.5cm,bb=0 0 1148 485]{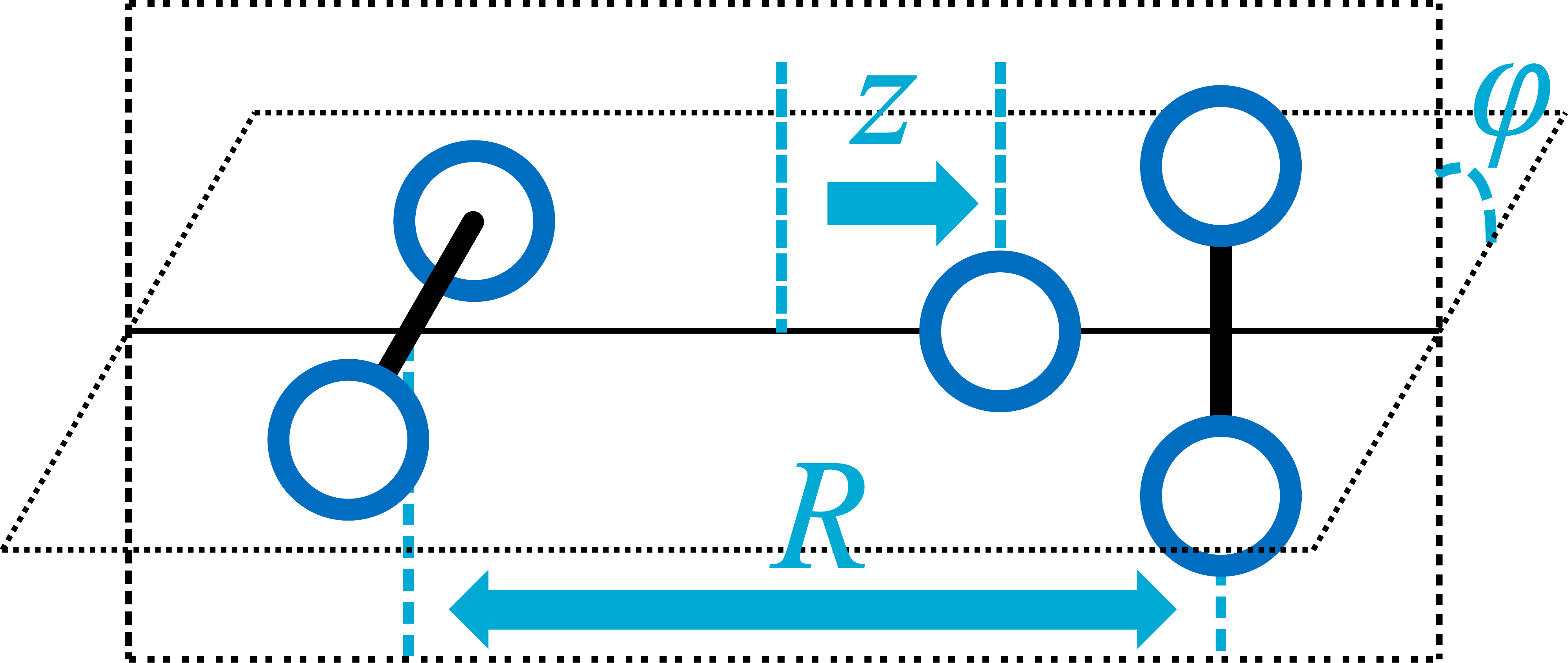}
          \caption{
          (color online).
          $\text{H}_3^+ + \text{H}_2 \rightarrow \text{H}_2+ \text{H}_3^+$ reaction can be written by three coordinate $\varphi,R,z$ depicted on picture.
          }
        \label{fig:fig3}
      \end{figure}
      
    In the present investigation, we treat the dynamics of $\mathrm{H}_5^+$
    by confining it into a three degrees-of-freedom system. 
    The dynamical variables are the center-of-mass distance $R$ between the two $\mathrm{H}_2$ moieties,
    the position $z$ of the central hydrogen atom along the center-of-mass axis, 
    and the torsional angle $\varphi$ of the two $\mathrm{H}_2$ as shown in Fig.~\ref{fig:fig3}.
    The coordinate $z$ corresponds to the proton exchange reaction between the two moieties, 
    while the angle $\varphi$ corresponds to the torsional isomerization. 
    We calculated the potential energy surface at the CCSD(T) level 
    which is the same level with the previous calculation \cite{Xie2005}.
    The \textit{ab initio} calculations were performed at 439 points 
    in the range $0\leq |z| \leq 0.4~ \text{\AA}$ and $2.09~ \text{\AA} \leq R \leq 2.51~ \text{\AA}$,
    with the $\mathrm{H}_2$ bond lengths optimized for each given value of $(z,R,\varphi)$.
    By checking the energy value, this region was confirmed to be sufficient to 
    describe the motion with total energy below 200 cm$^{-1}$.
    The potential energy values were then fitted to a cubic order polynomial in $(z^2,R,\cos 2 \varphi)$.
    The maximum fitting error was 0.8 cm$^{-1}$, sufficiently small considering 
    the total energy 170 cm$^{-1}$ of the trajectories run in the present investigation.
    The structures and energies of the four lowest stationary points 
    of the fitted surface are listed in Table~\ref{tab:eq} and compared with the literature values.\cite{Xie2005}
    The mathematical expression of the fitted potential energy surface
    is available as the supporting information to this article.
    
    \begin{table}
        \caption{
        Structures and energies of four lowest stationary points of $\mathrm{H}_5^+$.
        The energies are given relative to the first equilibrium point. 
        }\label{tab:eq}
      \begin{tabular}{ccccccl}
        \hline
         & $\varphi$ & $R$ / \AA & $z$ / \AA & Energy / cm$^{-1}$ & Ref.~\onlinecite{Xie2005} \cr
        \hline
        1 & $\pi/2$ & 2.18 & 0.19 & (ref.) & (ref.) & global minimum\cr
        2 & $\pi/2$ & 2.11 & 0 & 48.6 &  48.4 & index-one saddle\cr
        3 & 0 & 2.19 & 0.21 & 95.9 &  96.4 & index-one saddle \cr
        4 & 0 & 2.12 & 0 & 162.7 & 162.8 & index-two saddle \cr
        \hline
      \end{tabular}    \end{table}
    
    We use this three-dimensional system as an illustrative model to demonstrate the concepts
    introduced in Sec.~\ref{sec:th}.
    Note however that the real $\mathrm{H}_5^+$ system has larger DoFs
    (nine internal modes and three rotational modes). 
    Quantum effects must also be considered for the complete treatment of this system.
    We here briefly mention that the concept of reactivity boundaries around the index-one saddle point
    has recently been extended to incorporate ro-vibrational couplings\cite{NFrotSK,NFrotUC}
    and quantum effects\cite{QNFrev,QTDNF}. 
    It will be an important future work to combine these studies with the generalized reactivity boundaries
    proposed in the present paper.
    In the present numerical calculation we confine the system configuration 
    into the three-dimensional subspace mentioned above for the sake of simplicity.
    We still note the global minimum, the three lowest saddle points and their unstable directions
    are all included in this subspace, 
    while the motions transverse to this subspace are bath mode oscillations.
    This three-dimensional model is therefore expected to capture 
    some of the essential properties of the isomerization and the proton exchange processes
    in the real $\mathrm{H}_5^+$ system with low energies.

\begin{figure}
      \includegraphics[width=8.5cm,bb=0 0 230 230]{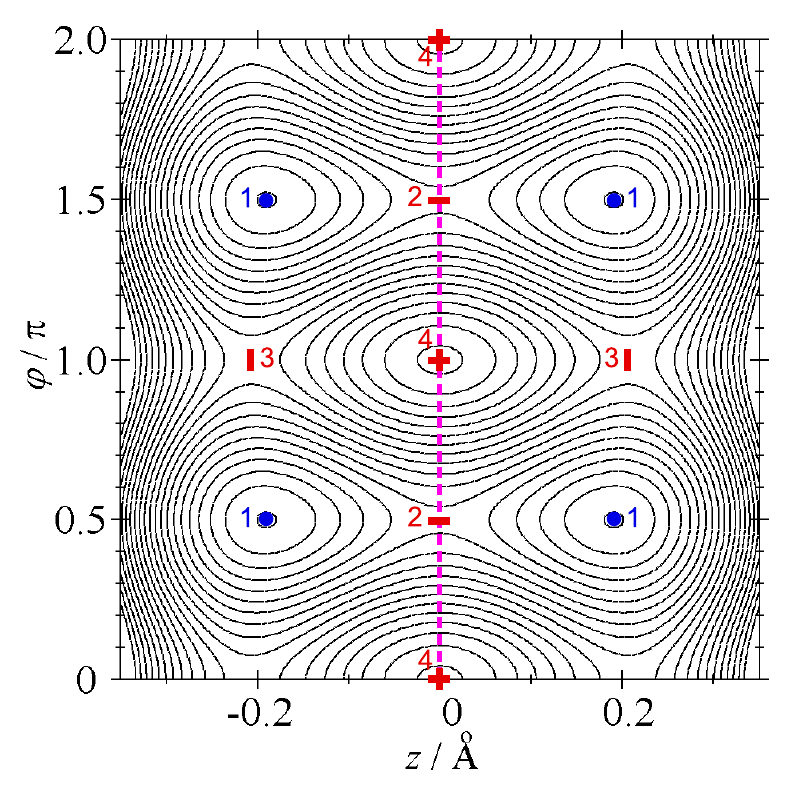}
        \caption{The potential energy surface as a function of $z$
        and $\varphi$, representing the proton exchange and the 
        torsional motion, where the other coordinate $R$
        is optimized at each point $(z,\varphi)$.
        Each number corresponds to each stationary point listed in
        Table \ref{tab:eq}.
        Blue points, red bars, and red cross denote the potential minima, 
        index-one saddles, and index-two saddle, respectively.
        Contours are spaced with 10 cm$^{-1}$. The initial condition 
        for the calculation of the reactivity boundaries
        are prepared at $z=0$, where index-one saddle
        points {\bf 2} and index-two saddle points {\bf 4} are located (pink dashed line).
        }
      \label{fig:fig4}
    \end{figure}

    There are two index-one saddle points, denoted as {\bf 2} and
    {\bf 3}, that correspond to
    the proton exchange and the torsional isomerization, respectively. 
    The highest of these four stationary points is an index-two saddle point,
    denoted as {\bf 4}, 
    representing a concerted reaction of the proton exchange and the torsion.
    Figure~\ref{fig:fig4} depicts the two-dimensional potential energy surface
    in $z$ and $\varphi$ where the $R$ is relaxed to the minimum energy for each 
    given value of $(z,\varphi)$.
    There are four symmetrically equivalent points corresponding to the global minimum {\bf 1}.
    Similarly the saddle points {\bf 2}, {\bf 3}, and {\bf 4} have two, four, and two equivalent 
    points, respectively.
    
    The dynamical calculations of the present three-dimensional model of $\mathrm{H}_5^+$
    are performed by integrating the equation of motion given by the following Hamiltonian
    \begin{align}
    H = \frac{1}{I_\varphi} {p_\varphi}^2 
    + \frac{1}{2\mu_R} {p_R}^2
    + \frac{1}{2\mu_z} {p_z}^2
    + V(\varphi,R,z),  
    \end{align}
    where $p_\varphi$ is the angular momentum conjugate to the torsional angle $\varphi$,
    and $p_R$ and $p_z$ are the linear momenta conjugate to $R$ and $z$, respectively.
    The reduced masses are 
    \begin{align}
    \mu_z = \frac{4}{5} m_\mathrm{H},
    && \mu_R = m_\mathrm{H},
    \end{align}
    where $m_\mathrm{H}$ is the mass of the hydrogen atom, 
    and $I_\varphi$ is the moment of inertia of H$_2$.
    
  \subsection{Reactivity boundary in $\text{H}_5^+$}
    
    As described in Sec.~\ref{ssec:RB}, the reactivity boundary typically consists of
    trajectories emanating from an invariant manifold in the intermediate
    region. 
    It is calculated by propagating the system, 
    either forward or backward in time, from the
    close vicinity of the invariant manifold. In the present investigation we
    focus on
    the proton exchange reaction from $\mathrm{H}_2 +
    \mathrm{H}_3^+$ to $\mathrm{H}_3^++\mathrm{H}_2$, to 
    demonstrate the extraction of reactivity boundary. 
    The configuration $\mathrm{H}_2 +
    \mathrm{H}_3^+$ corresponds to a region with $z>0$ and
    $\mathrm{H}_3^++\mathrm{H}_2$ with $z<0$. The
    intermediate region thus lies on some region around 
    $z=0$. In this case 
    the surface defined by $z=0$ and $p_z=0$ serves as an invariant
    manifold due to the symmetry of the system. 
    This means that, once the system stays on that surface, it does
    perpetually irrespective of what values the other variables
    take. This invariant manifold is unstable in that any
    infinitesimally small deviation from the surface of $z=0$ and
    $p_z=0$ makes the system depart from the surface and
    fall down into one of the four well regions shown in Fig.~\ref{fig:fig4}.
    Therefore the reactivity boundaries are stable and unstable manifolds of $z=0,p_z=0$ in this case.
    The extraction of reactivity boundary can be carried out as follows: we
    first uniformly sample phase space points
    $(p_z=0,p_R,p_\varphi,z=0,R,\varphi)$ at a given total energy
    in that invariant manifold (see also Appendix for details). Second, we give the system
    a small positive deviations in $p_z$, and propagate it forward in
    time (corresponding to the origin-dividing set o$_2$ in
    Fig.~\ref{fig:fig2}). Those trajectories correspond to the
    generalization of $\xi_1=0$ with positive $\eta_1$ to divide the origin
    of trajectories for normal mode approximation in Fig.~\ref{fig:fig1}. 
    Likewise, the propagation of the system backward in time results
    in trajectories that divide the destination of trajectories,
    corresponding to the set d$_2$ in Fig.~\ref{fig:fig2}
    (Compare also with $\eta_1=0$ with negative $\xi_1$ for normal mode
    Hamiltonian in Fig.~\ref{fig:fig1}). Note here again that the 
    generalization involves two essential differences from the normal mode
    picture: one is the
    generalization to nonlinear Hamiltonian systems in which normal mode
    approximation does not hold, and the other is that the invariant manifold
    from which reactivity boundaries emanate can be associated not only with a
    single saddle point but with multiple saddle points with different indices.

    \begin{figure}
      \includegraphics[width=8.5cm,bb=0 0 284 155]{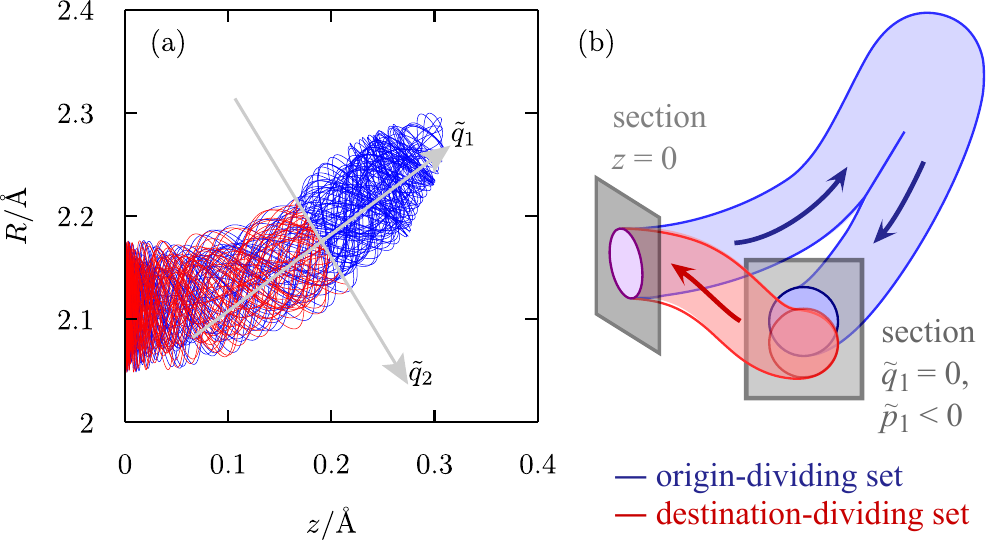}
        \caption{
        (color online).
        The reactivity boundaries of $\text{H}_3^+ +\text{H}_2 \rightarrow  \text{H}_2 + \text{H}_3^+$ reaction.
        (a) randomly sampled fifty trajectories from the
        destination-dividing set (red) and the origin-dividing set (blue), both constituting
        the reactivity boundaries, initiated from the section of $z=0$
        and $p_z \simeq 0$ projected on the $z-R$ space. The normal
        mode coordinates $\tilde{q}_1$ and $\tilde{q}_2$ at the potential
        minimum are shown by gray lines.
        (b) a schematic picture of reactivity boundaries depicted as
        ``tubes\cite{DeAlmeida1990}'' departing from $z=0$ and $p_z \simeq
        0$. Note here that the invariant manifold
        of $z=0$ and $p_z=0$ can involve multiple saddle
        points. 
        }
      \label{fig:fig5}
    \end{figure}
    \begin{figure*}
      \includegraphics[width=17cm,bb=0 0 502 378]{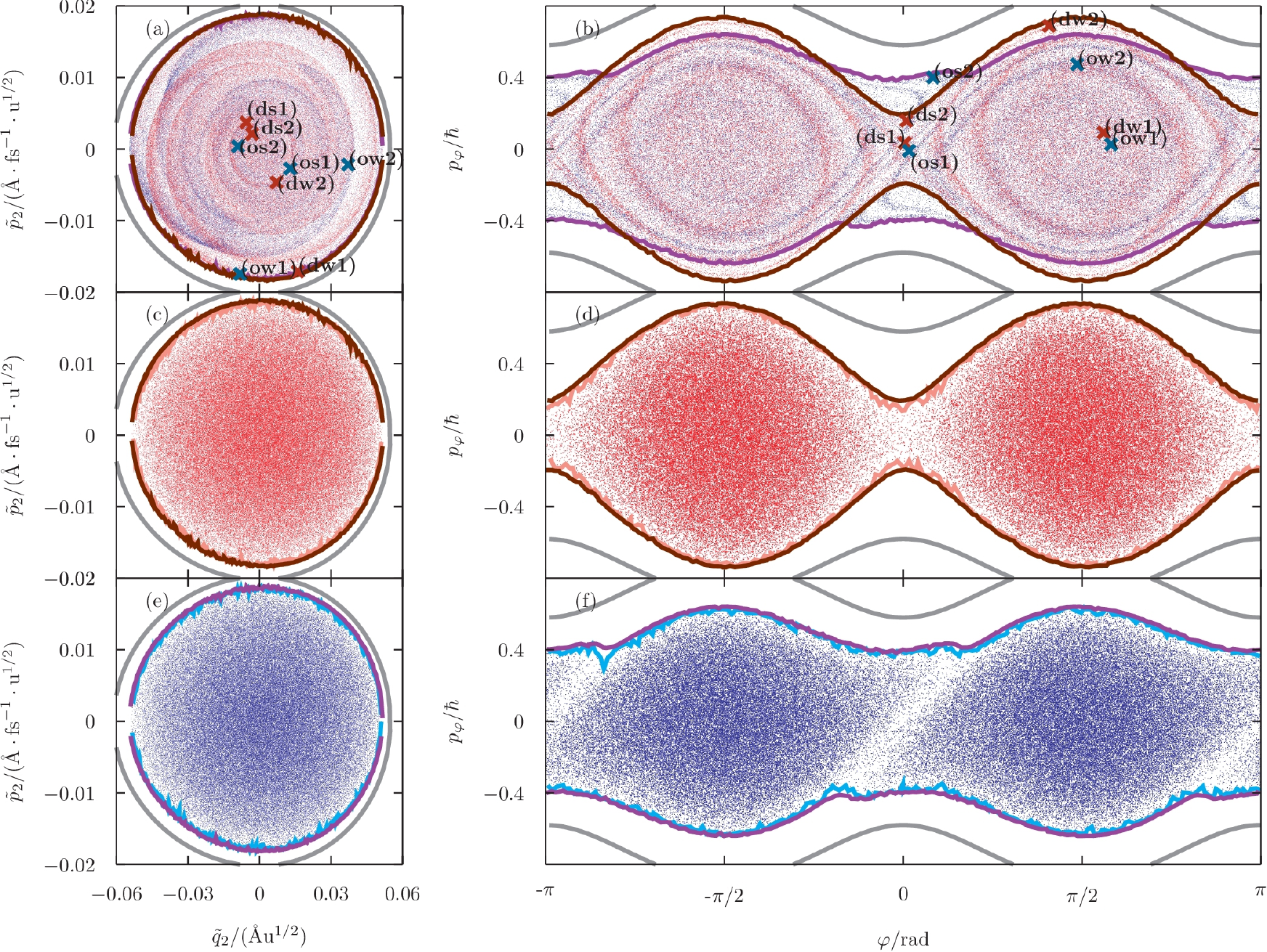}
        \caption{
        (color online).
        The reactivity of $\text{H}_3^+ +\text{H}_2 \rightarrow
        \text{H}_2 + \text{H}_3^+$ reaction on the section of $\tilde{q}_1=0, \tilde{p}_1<0$.
        100,000 trajectories are uniformly sampled on the surface
        of $z=0$ with positive
         momentum $p_z \simeq 0$ and evolved forward in time until
        they cross a surface defined by $\tilde{q}_1=0$ and
        $\tilde{p}_1 < 0$ by the normal mode coordinate $\tilde{\vect{q}}$
        and its conjugate momentum $\tilde{\vect{p}}$ at the potential minimum (see also Appendix for details). 
        The trajectories forming the origin-dividing set
        are shown by blue dots. Likewise, 100,000
        trajectories are similarly sampled on that surface with
        negative $p_z \simeq 0$ and propagated backward in time
        until they cross the surface. The trajectories forming
        the destination-dividing set
        are shown by red dots.
        (a) and (b): the projections of the first intersections of the
        destination-dividing set (red dots) and the origin-dividing set (blue dots) crossing the surface of $\tilde{q}_1=0$ and
        $\tilde{p}_1 < 0$ on the section, respectively, onto
        the $\tilde{q}_2$-$\tilde{p}_2$ space and the $\varphi$-$p_\varphi$
        space. The gray lines denote the boundaries of energetically
        inaccessible region. The values are defined by maximum and minimum of
        $p_\varphi$ at each $\varphi$. The cross symbols (dw1, dw2,...)
        represent the initial positions (on that place) of the trajectories 
        shown in \ref{fig:fig7}.
        (c), (d), (e), (f): the projections of 
        the phase space points that are going into the product side (red dots) and those that have come from the product side (blue dots)
        are depicted to conform ``inside'' of reactivity boundaries and 
        to check validity of the extraction of the reactivity boundaries.
        Orange lines in (c) and cyan   lines (e) are maximum/minimum $\tilde{p}_2$ of the sets of the reactive points.
        Brown lines  in (c) and purple lines (e) are maximum/minimum $\tilde{p}_2$ of the reactivity boundaries.
        Similarly,
        Orange lines in (d) and cyan   lines (f) are maximum/minimum $p_\varphi$ of the sets of the reactive points.
        Brown lines  in (d) and purple lines (f) are maximum/minimum $p_\varphi$ of the reactivity boundaries.
        }
      \label{fig:fig6}
    \end{figure*}

    Figure~\ref{fig:fig5} shows randomly chosen fifty samples from
    the origin-dividing set (blue) and the destination-dividing set (red) depicted on the $R$-$z$ space.
    The reactivity boundaries are only drawn until they first cross the section defined by $\tilde{q}_1=0$ and
    $\tilde{p}_1 < 0$ by the normal mode coordinate $\tilde{\vect{q}}$ and its conjugate momentum $\tilde{\vect{p}}$
    at the potential minimum (the normal mode coordinates are shown
    by the gray arrows in Fig.~\ref{fig:fig5} (a)). 
    The reactivity boundaries are four dimensional surfaces in
    an equienergy shell 
    which divide reactive and non-reactive trajectories as schematically
    shown in Fig.~\ref{fig:fig5}(b).
    Figures~\ref{fig:fig6}(a)(b) show the origin dividing set (blue)
    and the destination dividing set (red) on the $\tilde{q}_1=0, \tilde{p}_1 < 0$ section
    depicted by using 100,000 trajectories whose initial conditions are uniformly sampled on the $z=0,p_z\simeq 0$ section (see also Appendix for details). Let us look into how reaction selectivity existing in the 
    phase space can be rationalized or visualized in these projections.
    In Fig.~\ref{fig:fig6}(a), one can find few fingerprints of the
    reaction selectivity existing in the phase space with respect to
    the signs of the normal mode coordinate and momentum. 
    The reaction path is curved in the $R$-$z$ space as shown in
    Fig.~\ref{fig:fig5}(a) and the saddle points exist on the negative
    side of $\tilde{q}_1$. Because  Fig.~\ref{fig:fig6}(a) is the projection
    of the first intersections of the reaction boundaries across the surface
    of $\tilde{q}_1=0$ and $\tilde{p}_1<0$ (i.e., all dots on
    Fig.~\ref{fig:fig6}(a) are moving towards the surface of $z=0$), one may
    expect that $\tilde{q}_2<0$ or $\tilde{p}_2<0$ on that surface should
    enhance the reaction probability, resulting in a nonuniform distribution
    of the reaction boundaries on the $\tilde{p}_2$-$\tilde{q}_2$ space. 
    However, as seen in Fig.~\ref{fig:fig6}(a), the reaction boundaries 
    are distributed rather uniformly in this space (e.g., no preference in the
    sign of $\tilde{p}_2$). This implies
    that preparing $\tilde{q}_2<0$ or $\tilde{p}_2<0$ on that surface does
    not increase the ability of the system to climb the reaction barrier. 
    As shown in Fig.~\ref{fig:fig5}~(a), the trajectories oscillate rapidly
    in the $\tilde{q}_2$-direction and the bath mode coordinate
    change its sign many time before coming close to $z=0$,
    where the saddle points {\bf 2} and {\bf 4} for the proton transfer
    reaction are located, while they slowly adapt to the curved reaction
    pathway. The dynamics near the index-one and index-two saddle
    points, thus, seems not to be sensitive to the initial vibrational phase prepared in the well region.       

\begin{figure*}
      \includegraphics[width=17cm,bb=0 0 565 270]{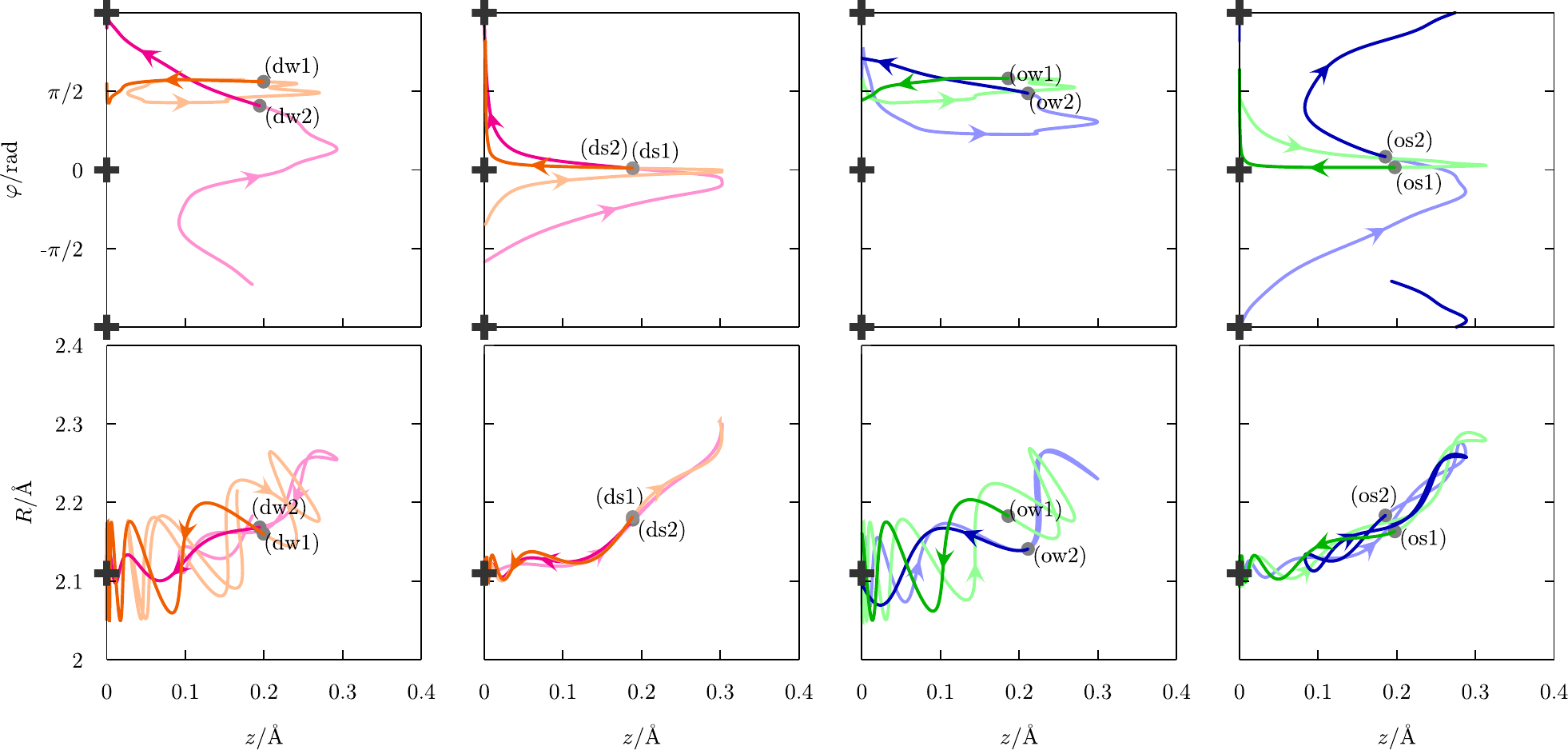}
        \caption{
        (color online). 
        The representative sample trajectories forming the reactivity boundaries
        in Fig. \ref{fig:fig6}(a) and (b) on the
        $\varphi$-$z$ space and the $R$-$z$ space.
        The gray points denote the locations in these spaces when those sample trajectories intersect the
        section of $\tilde{q}_1=0$ with $\tilde{p}_1 < 0$. The symbol {\bf $+$}
        denotes the location of the index-one point {\bf 2} or index-two
        saddle point {\bf 4}.  The difference of the location of the two saddle
        points is invisible in the $R$-$z$ projection with this resolution.
        The magenta and orange colored trajectories are of the
        destination dividing set. 
        The blue and green colored trajectories are of the origin dividing
        set. 
        The color grade represents the time course of trajectories obeying
        the Hamiltonian: time goes from the light to the dark grade, and the
        light and dark correspond to before
        and after the intersection of the section of $\tilde{q}_1=0$ with $\tilde{p}_1 < 0$. 
        For instance, trajectory (dw2) indicates that of the \underline{d}estination-dividing
        set in the \underline{w}ell region with `large' $|p_\varphi|$. Trajectory
        (os1) indicates that of the \underline{o}rigin-dividing
        set at the index-one \underline{s}addle region with `small'
        $|p_\varphi|$. 
        }
      \label{fig:fig7}
    \end{figure*}

    Next let us turn to the $p_\varphi$-$\varphi$ projection in Fig.~\ref{fig:fig6}(b). 
    The reaction boundaries, both the destination dividing set (red points
    in the figure) and the origin dividing set (blue), are confined in smaller
    values of $|p_\varphi|$ compared to energetically accessible
    values. This is because the energy is more distributed into the reactive
    mode when the momentum in the $\varphi$-direction is smaller. 
    In contrast to the $\tilde{p}_2$-$\tilde{q}_2$ space, the reaction selectivity 
    existing in the phase space manifests nonuniformity of the distribution
    of these reaction boundaries in the $p_\varphi$-$\varphi$ space. 
    The confinement of the destination-dividing set (red) in smaller
    $|p_\varphi|$ is more pronounced in $\varphi\approx0$ than in
    $\varphi\approx\pi/2$, while the range of $|p_\varphi|$ of the
    origin-dividing set (blue) is more uniform in $\varphi$. 
    Note that {$\varphi=0$} corresponds to the planar configurations that
    involve both the index-one
    saddle points {\bf 3} and the index two saddle points {\bf 4} (see
    Table~\ref{tab:eq}) and the reaction must proceed over the index-two
    saddle when $\varphi \approx 0$ (Fig.~\ref{fig:fig4}). 
    The relative barrier height through the index-two saddle {\bf 4} 
    for the proton transfer with $\varphi=0$ is $162.7-95.9=66.8\
    \mathrm{cm}^{-1}$ which is higher than
    the barrier height  through the index-one saddle {\bf 2} 
    with $\varphi=\pi/2$, 48.6 cm$^{-1}$ as seen from
    Table~\ref{tab:eq}. 
    In order to put sufficient energy into the reactive mode to overcome the
    barrier, therefore, the momentum $p_\varphi$ in the $\varphi$-direction
    must be confined into much smaller values $|p_\varphi|$ for
    $\varphi\approx0$ than for $\varphi\approx\pi/2$ due to the
    conservation of total energy of the system. 
    This interpretation, done by the relative barrier height
    with constant $\varphi$, is consistent with the plot of the
    sample trajectory (ds1) for small
    initial $|p_\varphi|$ in Fig.~\ref{fig:fig7}.
    shows some representative sample trajectories in the $\varphi$-$z$
    and $R$-$z$ spaces, whose locations in the $\tilde{p}_2$-$\tilde{q}_2$
    and the $p_\varphi$-$\varphi$ spaces are also indicated as symbols in
    Figs.~\ref{fig:fig6}(a)(b). It is seen that the
    motions along the reactive direction (approximately the
    $z$-direction) take place more rapidly than that along the
    $\varphi$-direction and the value of $\varphi$ does not change much
    during the course of the reaction. 

    On the other hand, the trajectories approaching to the surface of $z=0$ and $p_z<0$ 
    with large values of $|p_\varphi|$ at $\varphi\approx\pi/2$ at the section
    correspond to the motion starting from the well region 
    and approach to the index-two
    saddle {\bf 4}, as shown in the $z-\varphi$ plane in Fig.~\ref{fig:fig7} (dw2).
    This is contrasted with the trajectories starting with small $|p_\varphi|$ at $\varphi\approx\pi/2$
    and approaching to the index-one saddle {\bf 2} (dw1). 
    If we regard $(p_\varphi,\varphi)$ as roughly corresponding to the nonreactive mode,
    this situation seems to be counter-intuitive in that when the nonreactive
    degree of freedom is more 
    excited (i.e., larger 
    $|p_\varphi|$) the system is more likely to
    approach to the higher index-saddle with larger barrier height.
    This arises from the fact that the ``reaction direction'' for
    proton transfer through the index-two saddle is not simply
    along $z$ but runs diagonal in the $z$-$\varphi$ plane as the system
    goes from the well directly to the index-two saddle {\bf 4}. The
    large momentum $|p_\varphi|$ is also used for approaching to
    the higher barrier of the index-two saddle {\bf 4} 
    and, therefore, the large initial value $|p_\varphi|$ is 
    favored for the reaction over the index-two saddle. 
    All the above discussions explain the nonuniformity of the range of
    $p_\varphi$ with respect to $\varphi$ for the destination-dividing set
    (red) in Fig.~\ref{fig:fig6}~(b).

    Compared to the destination-dividing set, the origin-dividing set
    is more uniformly distributed along $\varphi$ (see blue dots in
    Fig.~\ref{fig:fig6}~(b)). This arises from the choice of the cross
    section for observing the reaction boundaries. We chose the section of
    $\tilde{q}_1=0$ with $\tilde{p}_1 < 0$ that is located at the
    potential minimum. With this choice, we are observing the
    origin-dividing set after it is bounced by the potential wall in
    the large-$z$ region (Fig.~\ref{fig:fig5}~(a)). As seen in the sample
    trajectories (os1),(os2),(ow1),(ow2) in Fig.~\ref{fig:fig7}, the
    value of $\varphi$ changes during the stay in the well region.
    The change of $p_\varphi$ due to the energy exchange between the $\varphi$-mode and the others 
    can also be seen by the direction of the trajectories.
    Therefore the longer time between the preparation (at $z=0$) and 
    the observation ($\tilde{q}_1=0$ with $\tilde{p}_1 < 0$) of the destination-dividing set
    than the origin-dividing set causes some further ``mixing'' in ($\varphi,p_\varphi)$ 
    and the reaction selectivity is lost compared
    to the direct cross section as observed for the destination-dividing set in Fig.~\ref{fig:fig6}.

    Reactivity boundaries are four dimensional objects, and
    we cannot capture their full characteristics by the two dimensional projections.
    In contrast to normal mode approximation or normal form theory
    locally expanded in the vicinity of a single saddle point, for our
    present general footing, the analytic formula of the
    underlying reaction coordinate is hard to derive and the invariant
    manifold locally extracted in the vicinity of a single point or a
    collection of multiple saddle points with different indices might not
    necessarily provide the boundary to divide the fates of the reactions
    originated from the well region far apart from the saddles.\cite{Nagahata2013a}
 
    To check 
    the validity of our numerical extraction of reactivity boundaries,
    we note the fact that both reactive and non-reactive trajectories must exist in the vicinity of the reactivity boundaries.
    We therefore check the reactivity of trajectories in the vicinity of each sampled point
    $(\tilde{p}_2,\tilde{q}_2,\varphi,p_\varphi)$ on the reactivity boundaries on the section.
    Sampling was made of phase space points $(\tilde{p}_2^\prime,\tilde{q}_2^\prime,\varphi^\prime,p_\varphi^\prime)$
    that satisfy
    \begin{eqnarray}
      \left|
      \frac{\tilde{p}_2^\prime-\tilde{p}_2}{0.02 \text{\AA} \mathrm{u^{1/2} fs^{-1}}}
      \right|^2
      +\left|
      \frac{\tilde{q}_2^\prime-\tilde{q}_2}{0.06 \text{\AA} \mathrm{u^{1/2}}}
      \right|^2&\\
      +\left|
      \frac{\varphi^\prime-\varphi}{\pi}
      \right|^2
      +\left|
      \frac{p_\varphi^\prime-p_\varphi}{0.8\hbar}
      \right|^2&
      =10^{-20}
    \end{eqnarray}
    for all the sampled points $(\tilde{p}_2,\tilde{q}_2,\varphi,p_\varphi)$ of the reactivity boundaries. 
    As expected, both reactive and non-reactive trajectories were found from this sampling (data not shown).
    
    To give more visual representation for
    the validity of our numerical extraction of reactivity boundaries, 
    we uniformly sampled 1,000,000 points on the $\tilde{q}_1=0,
    \tilde{p}_1 < 0$ section in the well region and
    propagated them forward and backward in time.
    The phase space points that turned out to go into the other well
    region in the forward time propagation are shown in (c) and (d) in
    Fig.~\ref{fig:fig6} by projection on the $\tilde{q}_2-\tilde{p}_2$
    space and the $\varphi-p_\varphi$ space.
    Those that turned out to have come from the other well 
    in the backward propagation are shown in (e) and (f).
    Of the total 1,000,000 sampled points, about 100,000 were found
    to be reactive trajectories.
    As can be seen in Fig.~\ref{fig:fig6}(c)-(f), a good coincidence was
    observed in the maximum/minimum $\tilde{p}_2$ and $p_\varphi$ at each
    $\tilde{q}_2$ and $\varphi$ between the reactive trajectories
    (corresponding to the inside of ``tubes\cite{DeAlmeida1990}'' in Fig.~\ref{fig:fig5}(b)) and the
    reactivity boundaries. In any neighborhood of
    the reactivity boundaries extracted from the surface of $z=0$ and $p_z
    \simeq 0$ apart from the well regions, reactive trajectories
    exist in the projected space.  
    The results, therefore, also give some support (necessary condition)
    to the validity of the
    reactivity boundaries calculated in the present investigation.

\section{Conclusion and Perspectives}
    In this article, the concept of reactivity boundary, which is an
    invariant manifold lying between reacting and non-reacting
    trajectories in the phase space, was revisited and generalized. It is
    defined as a set of trajectories that converge into a %n invariant
    %set
    seed of reactivity boundaries.
    The latter is located between the reactant and the product regions, 
    and goes neither into the reactant or the product, in either
    forward or backward time propagation. When only one saddle point controls the reaction
    dynamics and the energy is not very high above the saddle point, the
    reactivity boundaries are readily extracted analytically by normal
    form theory. The definition given here is, however, not limited to such
    cases but generalized to a single reaction passing through multiple saddle points including higher index saddles. 

    The reactivity boundaries constitute a
    skeleton of the phase space of the reaction
    system. Observation of their locations in certain cross sections tells which
    initial conditions can lead to chemical reactions. We applied the
    concept of the reactivity boundaries to the three-dimensional model system
    of the proton exchange reaction
    associated with a bottleneck made up of two index-one saddles ({\bf 2}) and two index-two saddles ({\bf 4})
    in $\mathrm{H}_5^+$ cation.
    The bath mode
    vibration represented by the normal mode $(\tilde{p}_2,\tilde{q}_2)$ was found to 
    be almost separate from the reactive mode, and the fast change of its vibrational phase 
    masked the reaction selectivity existing in the phase space.
 
    On the other hand, the reaction selectivity in the phase space
    manifested high degree of selectivity for the
    torsional motion, related to the existence of multiple types of saddle 
    points for different values of the torsion angle. In addition to the reaction
    through the index-one saddle {\bf 2} of the proton exchange, 
    two limiting behaviors of reacting trajectories were identified.
    In one group, the trajectories go from the index-one saddle {\bf 3} of the torsion isomerization
    to the index-two saddle {\bf 4}. 
    Small initial values of the torsional angular momentum $|p_\varphi|$ is favored
    for this reaction pathway because of the high energy difference between
    the index-two saddle point {\bf 4} and the index-one torsion
    saddle point {\bf 3}. The other group of the reacting
    trajectories is those going directly from the well region to the
    index-two saddle {\bf 4}. For this group, high initial values 
    of $|p_\varphi|$ is favored
    because the reaction pathway runs diagonal in the $z$-$\varphi$ plane 
    rather than parallel to the $z$-direction. These pictures of the
    reaction dynamics were obtained with the help of the concept of reactivity boundaries 
    stated in the present paper.

    In this article we have focused on the first intersection of the
    reactivity boundaries across the section of $\tilde{q}_1 = 0$ with
    $\tilde{p}_1 < 0$ located in the well region. This corresponds to 
    the fast stage of the reaction process, that is, ``before leaving from that
    well'' and ``after entering with one reflection back by the
    potential wall in that well.'' Reactivity boundaries
    also enable us to quantify the slow stage of the process by the
    projection of the second, third, fourth intersections of the
    boundaries onto, e.g., the $\varphi-p_\varphi$ space. Distributions
    of such intersections on some projected spaces can trace how statistical 
    properties may emerge for slower timescales (yielding a more uniform
    distribution), making conventional statistical rate theories applicable. 
    Note that as demonstrated in this article the first intersection
    corresponding to the reactive initial conditions are distributed in
    a nonuniform manner, to which conventional statistical rate theories
    are not applicable. The essential understanding of reactions requires reactivity
    boundaries that enable us to predict the fate of reactions independent of
    which timescale to be considered.

    In the extraction scheme of reaction boundary presented in
    Sec.~\ref{ssec:RB}, we have not restricted the definition of states to
    a local equilibrium state in which highly-developed chaos is implicitly postulated.
    As known, at least for two DoFs systems in
    Ref.~\onlinecite{Davis1986,MacKay1984}, there may exist several dynamic states within a
    single potential well whose number and the reaction rate constants among them are
    energy-dependent. The definition of states in Sec.~\ref{ssec:RB} can
    involve such nonergodic states. In addition, as discussed in the text,
    the seed of reactivity boundaries %invariant manifold 
    existing in between the states 
    involves not necessarily only one single saddle point but also several
    saddle points with different indices.
 
    The practical methods for extracting the reactivity boundaries, however,
    need still much to be considered. When only one saddle point plays a
    dominant role in determining the occurrence of the reaction, normal
    form theory readily extracts the seed of reactivity boundaries in the analytical
    way. In contrast, there is still no practical method applicable to
    general cases where more than one saddle points are involved in the
    reaction process. In the present investigation, because of the
    preknowledge concerning the existence of symmetry, we can 
    identify
    the %invariance of the manifold 
    seed of reactivity boundaries in the intermediate region. 
    When the symmetry cannot be exploited easily, it is still a
    challenging future work to devise convenient methods to extract
    %invariant manifolds. 
    seeds of reactivity boundaries\cite{Nagahata2013a}.

\section{Acknowledgment}
    TK has greatly benefited and been inspired from many discussions 
    with Prof. Oka and his enthusiasm on how new
    concepts emerge more than we may expect when two different
    disciplines meet with each other such as chemistry and astronomy in nature. 
    TK would like to dedicate this article using the concept of 
    chemistry and celestial mechanics to him in token of his gratitude
    for Prof. Oka's insightful thought.
    This work has been partially supported by JSPS, Research
    Center for Computational Science, Okazaki, Japan, Grant-in-Aid for
    Young Scientists (B) (to SK), Grant-in-Aid for challenging
    Exploratory Research (to TK), and Grant-in-Aid for Scientific
    Research (B) (to TK) from the Ministry of Education, Culture, Sports,
    Science and Technology.
  
  \bibliography{jcpa_nagahata}

\section{Appendix: Uniform sampling}
    Here we explain how we sample the uniform distributions under
    constraints to depict the reactivity boundaries and the sets
    of reacted/reacting trajectories, i.e., those having just
    crossed the surface of $z=0$ from the product well and those being
    about to cross the surface,
    in the reactant well described in Sec. IIIB. 
    To depict reactivity boundaries, we sample the position coordinate $(R,\varphi)$ according to the following distributions.
    \begin{eqnarray} \label{eq:ussd}
      &&\rho(R,\varphi;z=0,p_z=0,H=E) \nonumber\\
      &&\propto \int \delta(E-H(\mathbf{p},\mathbf{q}))\delta(z)\delta(p_z)dp_R dp_z dp_\varphi dz \nonumber\\
      &&\propto \sqrt{E-V(R,\varphi;z=0)}.
    \end{eqnarray}
    Here we define
    $\bar{\rho}_{\text{sd}}(R,\varphi)=\sqrt{\frac{E-V(R,\varphi;z=0)}{E-V_0}}$
    yielding
    $0<\bar{\rho}_{\text{sd}}<1$,
    where $V_0=\min_{R,\varphi}V(R,\varphi;z=0)$.
    We employ the rejection method \cite{Press2007} to
    sample phase space points with the distribution
    $\bar{\rho}_{\text{sd}}$.
    We first sample points uniformly in the range of
    $\varphi\in[-\pi,\pi]$ and $R\in[2\text{\AA},2.2\text{\AA}]$
    which include the whole energetically accessible region. 
    The point is accepted or rejected by the following criterion:
    \begin{equation}
      \begin{cases}
      \text{accept}  & \bar{\rho}_{\text{sd}}(R,\varphi)>\text{RAND},\\
      \text{reject} & \text{otherwise},
      \end{cases}
    \end{equation}
    where $\mathrm{RAND}$ is a uniform random number from $0$ to $1$.
    Then we perform sampling of the momentum for each sampled configuration as follows:
    \begin{eqnarray*}
      p_R=\sqrt{2(E-V)}\sin \theta/m_R,\\
      p_\varphi=\sqrt{2(E-V)}\cos \theta/(I_\varphi/2),
    \end{eqnarray*}
    where $\theta$ is a uniform random number from $-\pi$ to $\pi$.

    Similarly, to depict the sets of reacted/reacting trajectories in the reactant well, we sample phase space points according to the following distribution:
    \begin{eqnarray} \label{eq:uswl}
      &&\rho(\tilde{q}_2,\varphi;\tilde{q}_1=0,\tilde{p}_1<0,H=E) \nonumber\\
      &&\propto \int \delta(E-H(\mathbf{p},\mathbf{q}))\Theta(-\tilde{p}_1)
      \delta(\tilde{q}_1) d \tilde{p}_1 d \tilde{p}_2 d \tilde{p}_\varphi d\tilde{q}_1  \nonumber\\
      &&\propto E-V(\tilde{q}_2,\varphi;\tilde{q}_1=0).
    \end{eqnarray}
    Here $\Theta(x)$ is the Heaviside step function, and we define
    $\bar{\rho}_{\text{wl}}(\tilde{q}_2,\varphi)=(E-V(\tilde{q}_2,\varphi;\tilde{q}_1=0))/(E-V_0)$,
    yielding
    $0<\bar{\rho}_{\text{wl}}<1$,
    where $V_0=\min_{\tilde{q}_2,\varphi}V(\tilde{q}_2,\varphi;\tilde{q}_1=0)$.
    We sample points uniformly in the range of 
    $\varphi\in[-\pi,\pi]$ and $\tilde{q}_2\in[-0.15\text{\AA}\mathrm{u^1/2},0.15\text{\AA}\mathrm{u^1/2}]$
    which include the whole energetically accessible region on this section. 
    We apply the same rejection method  \cite{Press2007} to construct $\bar{\rho}_{\text{wl}}$ distribution
    \begin{equation}
      \begin{cases}
      \text{accept}  & \bar{\rho}_{\text{wl}}(\tilde{q}_2,\varphi)>\text{RAND},\\
      \text{reject} & \text{otherwise}.
      \end{cases}
    \end{equation}
    Then we perform sampling of the momentum for each sampled configuration as follows:
    \begin{equation}
      \begin{cases}
      \text{accept}  & \sin\theta_1>\text{RAND} ,\\
      \text{reject} & \text{otherwise} .
      \end{cases}
    \end{equation}
    \begin{eqnarray*}
      \tilde{p}_\varphi&=&\sqrt{2(E-V)}\sin \theta_1 \sin \theta_2,\\ 
      \tilde{p}_1&=&-\sqrt{2(E-V)}\cos \theta_1, \\
      \tilde{p}_2&=& \sqrt{2(E-V)}\sin \theta_1 \cos \theta_2,
    \end{eqnarray*}
    since coordinate transformation to polar coordinates introduces phase space Jacobian $J=2(E-V)\sin \theta_1$,
    where $\theta_1,\theta_2$ are uniform random numbers from $0$ to $\pi/2$ and from $-\pi$ to $\pi$, respectively.

\end{document}